# Assessment of SFSDP Cooperative Localization Algorithm for WLAN Environment

Ebtesam Almazrouei, Nazar Ali, and Saleh Al-Araji
Khalifa University, Abu Dhabi, UAE
{Ebtesam.almazrouei, ntali}@kustar.ac.ae

*Abstract*—Cooperative localization for indoor WiFi networks have received little attention thus far. Many cooperative location algorithms exist for Wireless Sensor Network Applications but their suitability for WiFi based networks has not been studied. In this paper the performance of the Sparse Finite Semi Definite Program (SFSDP) has been examined using real measurements data and under different indoor conditions. Effects of other network parameters such as varying number of anchors and blind nodes are also included.

## I. INTRODUCTION

The main challenge facing Wi-Fi Cooperative Localization in indoor/urban environments is the multipath and non-line of sight problems that can degrade RSS and TOA based distance estimation techniques. The second major challenge is the design and development of robust algorithms to combine accurate range/distance measurements to localize APs in a network through centralized or distributed cooperative localization algorithms.

Cooperative localization for wireless sensor networks research has been vigorous over the last decade[1]. In the literature, there are many cooperative algorithms developed to locate a number of blind nodes (unknown position) with a number of anchors (known position) in wireless sensor networks (WSN). However, the suitability of these WSN algorithms for cooperative localizations in Wi-Fi based networks and in presence of multipath effects has not received similar attention. In [2, 3] it was concluded that centralized algorithms such as the Semi-Definite Programming (SDP) provide more accurate results than the distributed algorithms with similar cost [4]. In [5] Doherty estimated node positions based on connectivity-induced constraints in a sensor network. He solved the localization problem as a convex optimization (linear) using SDP. Ouyang, et al. [6] solved localization problem as a minimal optimization via SDP which obtained high accuracy with more complexity. Biswas and Ye [7] proposed a Finite Semidefinite Program (FSDP) algorithm to compute the approximate location of sensor with an accurate solution. Kim [4] developed the Sparse Finite Semidefinite Program (SFSDP) to solve large sensor network problem which can handle up to 6000 sensors in 2-dimensional problem.

In this work, the performance of existing centralized Cooperative Localization algorithms such as SFSDP is studied in the context of TOA based WiFi networks in indoor environments. The performance of the SFSDP algorithm developed by Kim [4] for sensor network is examined under realistic propagation channels that suffer from multipath and NLOS impairments. The empirical model for TOA ranging which was developed by the authors [2] is used to study the behavior of the SFSDP algorithm.

## II. WIRELESS NETWORK LOCALIZATION MODEL

For a specific network with blind nodes $s_i$ and anchors $a_r$, the Euclidean distances between the $i^{th}$ and $j^{th}$ blind nodes $d_{ij}$ and between the $i^{th}$ blind node and $r^{th}$ anchor $d_{ir}$ should be determined.

There will be a set of distance pairs $N_s$ which includes all the Euclidean distances between blind nodes such that $d_{ij}$ is not greater than the radio range $\rho$, where $N_s \in N_s^\rho$. The radio range $\rho$ is the signal maximum travel distance, i.e when the signal power equals the noise floor. Also, the Euclidean distance $d_{ir}$ such that $d_{ir} \in N_a$ and $N_a$ is a subset of $N_a^\rho$.

$$N_s^\rho = \{(i,j): 1 \leq i < j \leq m, \|s_i - s_j\| \leq \rho\} \quad (1)$$

$$N_a^\rho = \{(i,r): 1 \leq i \leq m, m+1 \leq r \leq n, \|s_i - a_r\| \leq \rho\} \quad (2)$$

where $s_i = [x_i, y_i]^T$ is the location of the $i^{th}$ blind node, $s_i \in R^\ell$, $i = 1,..., m$ in $\ell$ dimensional space and $m$ blind nodes. In this work $\ell = 2$ as only two dimensional networks are considered. Also, $a_r = [x_r, y_r]^T$ is the known location of anchor node $r$, $a_r \in R^\ell$, $r = m+1,..., n$.

The estimated distances $\hat{d}_{ij}$ and $\hat{d}_{ir}$ are computed as quadratic equations to be applied in SDP:

$$\|s_i - s_j\|^2 = (d_{ij})^2, \forall (i,j) \in N_s \quad (3)$$

$$\|s_i - s_r\|^2 = (d_{ir})^2, \forall (i,j) \in N_a \quad (4)$$

To solve the system of equations defined by the network problem using SDP, the minimization of the objective function is given as [8]

$$minimize \sum_{(i,j) \in N_x} \varepsilon_{ij}^2 + \sum_{(i,r) \in N_a} \varepsilon_{ir}^2 \quad (5)$$

$$\varepsilon_{ij}^2 = \left((\hat{d}_{ij})^2 - \|s_i - s_j\|^2\right)^2 \quad (6)$$

$$\varepsilon_{ir}^2 = \left((\hat{d}_{ir})^2 - \|s_i - s_r\|^2\right)^2 \quad (7)$$

where $\varepsilon$ is the error in distance estimation. The error in position estimation of blind nodes is the result of AWGN (with zero mean)presence and the propagation condition. For LOS channels, TOA estimation using WiFi Systems (20 MHz bandwidth) can be significantly corrupted by the multipath while in NLOS, both multipath and NLOS bias are involved. Therefore, the distance estimation model can be expressed as,

$$\hat{d}_{ij} = \|s_i - s_j\| + \varepsilon_{AWGN} + \varepsilon_{los/nlos} \quad (8)$$

$$\hat{d}_{ir} = \|s_i - a_r\| + \varepsilon_{AWGN} + \varepsilon_{los/nlos} \quad (9)$$

The authors in [2] modelled the TOA estimation error for LOS and NLOS propagation for stationary scenarios and in presence of AWGN with a variance $\sigma^2$ as a normal distribution with a mean and variance. The models were derived from real-time measurements carried out in an indoor environment [2, 9]. In this work, these models are incorporated in the SFSDP algorithm provided by Kim [4] in order to provide a realistic and practical evaluation of the algorithm's performance. To the best of the authors' knowledge, this is the first work that investigates the impact of realistic channel propagation conditions on the performance of a centralized cooperative localization algorithm.

In order to analyze the performance of the algorithms, the average position error $P_m$ is calculated for all blind nodes,

$$P_m = \left(\frac{1}{m}\sum_{i=1}^{m}\|\hat{s}_i - s_i\|\right) \quad (10)$$

## III. Simulation Setup and Analysis

A 30 m x 30 m 2-Dimensional network is generated using the SFSDP software with randomly placed anchors and blind nodes. The size is chosen based on the regular size of a typical WiFi network in indoor environments with a radio range $\rho$ = 15m. Experimental data are used for propagation errors for LOS and NLOS in stationary scenarios have means and variances of 6.98 m, 1.87 m for LOS and 16.06 m, 0.68 m for NLOS, respectively [2].

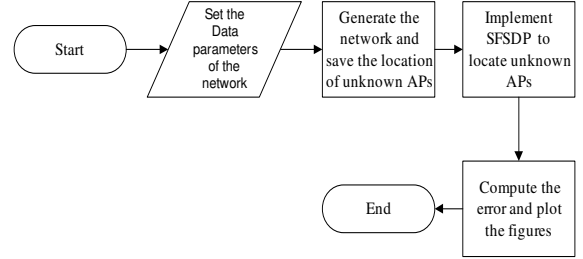

Fig. 1. The structure of network simulation methodology.

Fig. 1 illustrates the main steps of the system methodology. First, the network parameters are identified. Then, the software generates a 2-Dimensional network with a specified number of anchors and blind nodes, and save their positions. SFSDP is then implemented to solve the blind nodes position using the modified SFSDP with the noise model. After that, the position error $P_m$ is computed for each blind node and plots are generated of the true position and the computed ones.

### A. AWGN Noise and Multipath Effects

A 2-Dimensional network in an indoor environment with 50 blind nodes and 10 anchors are placed. The radio range $\rho$ is set to 15m. The nodes are placed randomly and three WiFi environment scenarios were considered; (1) with no additive noise and multipath effects (ideal channel), (2) with AWGN noise of a variance $\sigma^2 = 0.3$ m$^2$, and (3) with AWGN (0.3 m$^2$ variance) and LOS/NLOS multipath effects. The percentage of NLOS blind nodes is 50% (25 nodes) of the total blind nodes in the network.

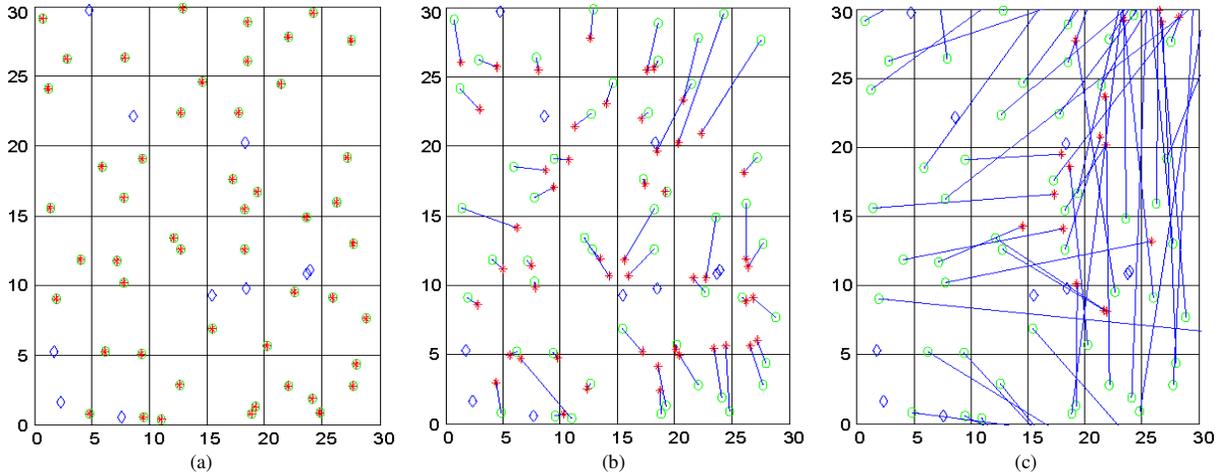

Fig. 2. The performance of SFSDP with (a) Ideal channel, (b) AWGN noise, and (c) LOS/NLOS + noise effects. Green circles are the true position of blind nodes, red stars are the computed position, blue diamonds are the anchor's position, and blue lines are the position errors.

The performance of the SFSDP algorithm in locating blind nodes within the 2-dimention (30 m x 30 m) WiFi network under the three scenarios is shown in Fig. 2. The blue diamonds are the positions of anchors, the green circles refer to the true location of blind nodes, and the red stars are the computed locations by the SFSDP for the blind nodes. The difference between the estimated and the original position of the blind nodes is indicated by the blue solid lines. From the figure, it is clear that the first scenario, with no noise or multipath effects, gives best results and the computed locations by SFSDP are exactly the same as the true locations of blind nodes.

Table 1 summarises some position errors $P_m$. The addition of measurement noise can cause a 2.48 m position error while the position error is increased dramatically to 23.23 m when realistic propagation error are introduced.

Table 1. Performance of SFSDP with no noise, noise, and NLOS Effect.

| Scenarios | Position error $P_m$ |
|---|---|
| 1. Ideal Channel | 9.5e-7m |
| 2. Measurement noise | 2.48m |
| 3. Measurement noise and propagation error | 23.23m |

### B. Effect of Number of Anchors

In this section, the effect of varying the number of anchors in the network on the SFSDP performance is explored. The number of anchors is increased gradually from 3 until it reached 50% of the total number of blind nodes (25) anchors while other parameters such as the radio range $\rho$, network dimensions and number of blind nodes ($m = 50$), are kept fixed. Fig. 3 shows the mean position error ($P_\mu$) defined in (11) for the three scenarios with $L = 100$ random trials while table 2 summarises these results. In each trial the nodes were positioned randomly by the program.

$$P_\mu = \frac{1}{L}\sum_{1}^{L} P_m \quad (11)$$

As expected, the SFSDP estimates the positions of blind nodes precisely for the ideal channel, i.e. scenario (1). When measurement noise ($\sigma^2 = 0.3$ m$^2$) is introduced, however, the position error $P_\mu$ follows an inverse relationship with the number of anchors; $P_\mu$ is 2 m, 1 m, and 0.9 m for 5, 15, and 25 anchors, respectively. A slightly different behaviour was exhibited under scenario (3), with LOS/NLOS multipath effects and noise, as the position error $P_\mu$ shows virtually no change with the number of anchors. This can be attributed to the overwhelming effect of multipath components on location estimation.

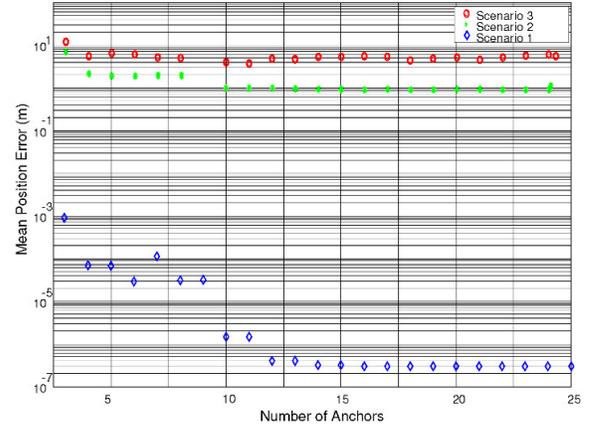

Fig. 3. 100 trial mean positon error vs number of anchors for the three scenarios.

Table 2. Effect of number of anchors on position error $P_\mu$.

| Scenarios | 1 | 2 | 3 |
|---|---|---|---|
| 5 Anchors | 6e-5m | 2m | 6m |
| 15 Anchors | 3.5e-7m | 1m | 6m |
| 25 Anchors | 3e-7m | 0.9m | 6m |

### C. The Effect of Density of the Network.

The effect of changing the number of blind nodes in a WiFi network is studied in this section. The number of anchors is always 30% of the total number of blind nodes while all other parameters are kept the same. Fig. 4 depicts the effect of increasing the density of the network (number of blind nodes) on the position error $P_m$ for 100 random trials for the three scenarios. Under scenario 1, there is a perfect match between the real and the estimated locationsof the blind nodes. Nevertheless, for scenarios 2 and 3, the algrithm seems undeciasive as there is fluctuations in the mean position error $P_\mu$. It concludes that the SFSDP algorithm fails to determine the position of the blind nodes when their number is varied.

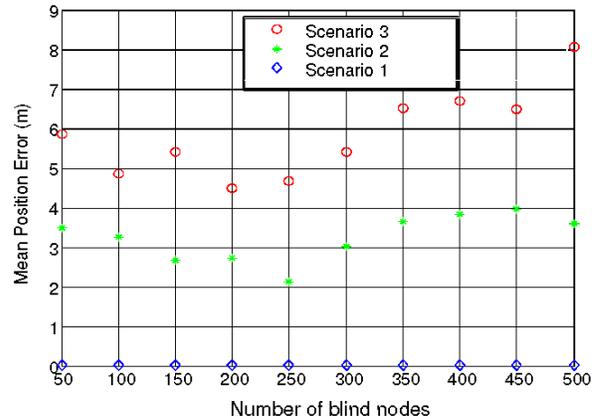

Fig. 4. Effect of network density on the mean positioning error for the three scenarios.

Table 3 also shows that the results of SFSDP under scenario 3 has the highest variance in position estimation.

Table 3. Performance of SFSDP with respect to network density.

| Scenarios | Position mean $P_\mu$(m) | |
|---|---|---|
| | Mean | Variance |
| Ideal Channel | 8.1264e-7 | 1.2536e-12 |
| Measurement Noise | 3.2304 | 0.3502 |
| Multipath and Measurement Noise | 5.8446 | 1.2303 |

*D. The Effect of Radio Range and NLOS Percentage*

The radio range $\rho$ for each blind node is important because it identifies its coverage area in the network. The position of blind nodes is estimated when it falls in the coverage area of the anchor or blind nodes. Also, the percentage of NLOS or multipath severity in the network impacts the localization accuracy of each blind node. In this section, the effect of multipaths is examined for three radio ranges. The radio range is varied from 15 m, 20 m, and 25 m, based on the size of the network being 30m x 30m. The number of anchors is kept at 3 while the percentage of NLOS range measurements is increased from 0 to 100%. Fig.5 illustrates the position error $P_m$ for 100 random trials as a function of NLOS percentage for the three different radio ranges. As expected, the position error $P_\mu$ increases with increasing NLOS effects. The 0% NLOS means the error is solely the result of noise.

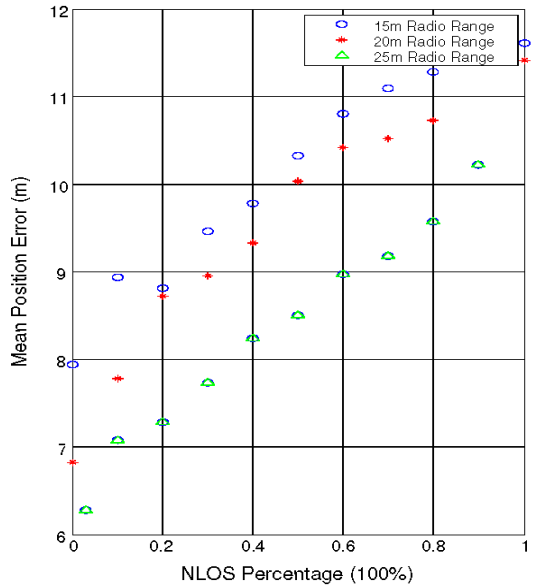

Fig.5.The effect of the NLOS percentage in the mean positioning error for different radio ranges.

IV. CONCLUSION

In this work, the performance of the SFSDP cooperative algorithm initially developed for Wireless Sensor Networks, has been examined for indoor and WiFi suitability. Empirical TOA ranging models, based on real time measurements and data, were developed by the authors [2] and incorporated in the SFSDP under different indoor conditions. This included, ideal channel, AWGN noise and a combination of noise and (LOS/NLOS) multipath effects. The results show that: 1) The performance of SFSDP in WiFi network under NLOS and propagation condition degrades, 2) Changing the WiFi network parameters such as number of blind nodes and number of anchors, radio range and NLOS percentages affect the accuracy of SFSDP in estimating the position of blind nodes. This work confirmed that the performance of the SFSDP localization algorithm for wireless sensor network has been drastically compromised when used in WiFi-based networks with real TOA propagation models. Therefore, there is a need to develop a more accurate cooperative localization algorithms for WLAN indoor applications.